\documentclass[12pt]{iopart}
\usepackage{epsfig}
\usepackage{iopams}
\eqnobysec
\newcommand{\be}{\begin{equation}}
\newcommand{\ee}{\end{equation}}

\newcommand{\ba}{\begin{array}}
\newcommand{\ea}{\end{array}}
\newcommand{\bea}{\begin{eqnarray}}
\newcommand{\eea}{\end{eqnarray}}

  \newcommand{\lab}{\label}

\newcommand{\nn}{\nonumber \\}

\newcommand{\bel}{\begin{equation}\label}

\begin{document}

\title{Supersymmetric Quantum Mechanics with Reflections}
\author{Sarah Post$^1$, Luc Vinet$^1$ and Alexei Zhedanov $^2$  }
\address{$^1$ Centre de Recherches Math\'ematiques. Universit\'e de Montr\'eal. Montr\'eal CP6128 (QC) H3C 3J7, Canada}
\address{$^2$ Donetsk Institute for Physics and Technology. Donetsk 83114, Ukraine}
\ead{post@crm.umontreal.ca, luc.vinet@umontreal.ca, zhedanov@fti.dn.ua}
\begin{abstract} We consider a realization of supersymmetric quantum mechanics where supercharges are differential-difference operators with reflections. A supersymmetric system with an extended Scarf I potential is presented and analyzed. Its eigenfunctions are given in terms of little $-1$ Jacobi polynomials which obey an eigenvalue equation of Dunkl type and arise as a $q\rightarrow -1$ limit of the little $q-$Jacobi polynomials. Intertwining operators connecting the wave functions of  extended Scarf I potentials  with different parameters are presented. 
\\
{}\\
Keywords: supersymmetric quantum mechanics; Dunkl and reflection operators; Scarf  potential; little-1 Jacobi polynomials
\end{abstract}
\pacs{ 02.30.Gp, 02.30.Hq,  03.65.Fd, 03.65.Ge,  12.60.Jv }
\ams{15A18, 05E35, 33D45, 34Kxx, 81Q60 }
%\maketitle
\section{Introduction}
Since its introduction by Witten \cite{1}, 30 years ago, supersymmetric quantum mechanics has been widely developed and has found numerous applications, both physical and mathematical. Among the many reviews and book published on this topic, the reader might consult  \cite{2} for background relevant to this paper. 

We consider here a realization of supersymmetric quantum mechanics that relies on the use of
reflection operators \cite{P1, P2}. Unlike the most standard approach it does not (necessarily) involve the presence of "spin-like" degrees of freedom and associated finite dimensional vector spaces, it implies however the presence of reflection operators in the Hamiltonians. We shall consider the simplest context of one-dimensional quantum mechanics. 

Hamiltonians with reflection operators have most notably arisen in the context of quantum many-body integrable systems of Calogero-Sutherland type \cite{3,4} and their generalizations with internal degrees of freedom \cite{5}. In these models, the constants of motion (and hence the Hamiltonians) are best expressed, and shown to be in involution, using exchange operators \cite{6,7,8, BDF, FLV} known in the mathematical literature as Dunkl operators \cite{9}. These are differential-difference operators that involve reflections. Such operators are needed to describe parabosonic oscillators \cite{10,11, 12}. Associated are deformed Heisenberg algebras that were  used in \cite{P1,P2} to design a supersymmetry without fermions
of the kind that will be of interest here. The exchange formalism has also proved  instrumental in demonstrating \cite{13} the superintegrability of certain models in the plane \cite{14}. Recently, symmetry algebras with reflection operators as elements have been examined, in the framework of a finite oscillator model \cite{15} or as the $q\rightarrow -1$ limit of the quantum algebra $sl_q(2)$ \cite{16}. Also, in the design of spin chains for quantum information transport, the property of mirror symmetry \cite{17} that is required for perfect transmission brings in equations involving reflection operators. While these studies provide many reasons to examine Hamiltonians with reflection operators, there is also intrinsic merit in the identification of exactly solvable quantum mechanical problems where supersymmetry manifests itself.

In mathematics, the Dunkl operators are central to the theory of multivariate orthogonal polynomials \cite{18} and there is currently much activity in the area of Dunkl harmonic analysis \cite{19}. Recently, two of us have authored and co-authored a series of papers \cite{20,21, 22, 23} showing that the set of classical orthogonal polynomials in one variable can be significantly enlarged by studying polynomial eigenfunctions of first-order differential operators of Dunkl-type. The simplest of these heretofore "missing" classical orthogonal polynomials are called little $-1$ Jacobi polynomials and will intervene below. As their name indicates, they can be obtained \cite{20} as a $q\rightarrow -1$ limit of the little $q$-Jacobi polynomials \cite{24}. 

The outline of this article is as follows. In section 2 we shall indicate in general terms, how supersymmetric Hamiltonians can be derived from Hermitian supercharges involving the reflection operator. The difference with the standard approach will be pointed out. We shall examine in section 3, the very simple case of a supersymmetric oscillator Hamiltonian with reflection. A more elaborate example will be provided in section 4, where an extension of the Scarf I potential \cite{25} will be introduced and studied. The eigenfunctions associated to this extended potential will be given in terms of little $-1$ Jacobi polynomials. The normalization is determined in \ref{AppendixA}.  Furthermore, intertwining operators connecting the wave functions of the supersymmetric Scarf I potentials with different parameters will be presented. A brief conclusion will follow. In \ref{AppendixB}, we provide examples of one-dimensional quantum Hamiltonians with reflection operator. They are not supersymmetric but their wave functions involve the generalized Gegenbauer
polynomials \cite{Chih, Belm} that share with the little -1 Jacobi polynomials the Dunkl-classical property\cite{20, Cheikh}.

\section{Supersymmetric Quantum Mechanics with Dunkl Supercharges}

To facilitate the comparison between the usual supersymmetric quantum mechanics and the one with reflections, let us first recall the basics of the standard approach. Let $H$ be a Hamiltonian; it is said to be supersymmetric if there are supercharges $Q$, $Q^\dagger$ such that the superalgebra relations 
\be H=\lbrace Q, Q^\dagger \rbrace, \qquad [Q, H]=0, \qquad [Q^\dagger, H]=0\ee
are realized. As usual, $\{ A, B \} =AB +BA$, $[A,B] =AB-BA.$ In the most simple setting of one-dimensional quantum mechanics, this is achieved by taking 
\be \label{Q} Q=\frac{1}{\sqrt{2}}(p-iW)b\ee
where $p=-id/dx$, $W=W(x)$ is the superpotential and $b,$ $b^\dagger$ are fermionic annihilation and creation operators satisfying 
\be b^2=(b^\dagger)^2=0, \qquad \{ b, b^\dagger\}=1, \ee
and represented by the $2\times 2$ matrices: 
\be b=\left[ \ba{cc} 0 &1\\ 0 &0\ea \right], \qquad b^\dagger=\left[ \ba{cc} 0 &0\\ 1 &0\ea \right].\ee
Upon calculating $\{ Q, Q^\dagger\}$ with $Q$ given by \eref{Q}, we readily find 
\be\label{25} H=\{ Q, Q^\dagger\}=\frac{1}2(p^2+W^2)+\frac12\frac{dW}{dx}\sigma_3\ee
where 
\be \sigma_3=[b, b^\dagger]=\left[\ba{cc}1& 0\\0&-1\ea\right].\ee
\Eref{25} gives the form of a supersymmetric Hamiltonian in one-dimension that has 2 supercharges $Q$ and $Q^\dagger$. One speaks of $N=1$ supersymmetry. Note that there are systems with only one Hermitian supercharge $Q=Q^\dagger$ such that $H=Q^2$. The Pauli Hamiltonian in the presence of a magnetic monopole is one such system \cite{26}. One speaks of $N=\frac12$ supersymmetry. In the following, we shall consider mostly such $N=\frac12$ (or chiral) supersymmetric problems. 

For reference, let us record the specific form of $H$ when 
\be\label{27} W=\frac{-\beta}{2\cos x}, \qquad -\frac{\pi}{2} \leq x \leq \frac{\pi}2, \ee
\bea H&=&-\frac{1}2\frac{d^2}{dx^2}+\frac{\left(\frac{\beta}{2}\right)^2}{2\cos^2 x}-\frac{\left(\frac{\beta}{2}\right)\sin x}{2\cos^2 x}\sigma_3\nn
\label{28}&=& \left[ \ba{cc} H_\beta &0 \\ 0&H_{-\beta}\ea \right]\eea
with 
\bel{29} H_\beta=-\frac{1}2\frac{d^2}{dx^2}+\frac{\beta(\frac{\beta}{2}-\sin x)}{4\cos^2 x}. \ee
This is a supersymmetrization of the Scarf I potential. Note that a more general 2-parameter form can be obtained \cite{2} by using $W=A\tan(\alpha x)-B\sec(\alpha x).$ This supersymmetric system has recently been further generalized in \cite{Quesne2009} using exceptional polynomials. 

Let us now indicate how the relation $H=Q^2$ can be realized by introducing reflections instead of "spin" degrees of freedom. Let $R$ denote the reflection operator:
\bel{210} \label{R} Rf(x)=f(-x).\ee
A  realization of supersymmetric quantum mechanics is obtained by taking as supercharge the following differential-difference operator of Dunkl type: 
\be\label{211} Q=\frac{1}{\sqrt{2}}\left( \frac{d}{dx} +U(x)\right)R+\frac{1}{\sqrt{2}}V(x),\ee
where $U(x)$ is an even function and $V(x)$ an odd function, 
\bel{212} U(-x)=U(x), \qquad V(-x)=-V(x). \ee
Since $R$ is symmetric $R^\dagger=R$, we easily see that $Q$ shares that property $Q^\dagger=Q.$ It is again a simple calculation to evaluate $Q^2$ and to find the following form for a supersymmetric Hamiltonian $H$:
\be \label{213} H=Q^2=-\frac12\frac{d^2}{dx^2}+\frac12(U^2+V^2)+\frac12 \frac{d U}{dx}-\frac12 \frac{dV}{dx}R. \ee
In this realization, unless $V$ is a constant, the operator $R$ appears in the Hamiltonian. It is of course possible to write \eref{213} in a $2\times 2$ matrix form. Consider to that end the Schr\"odinger equation $H \Psi=E\Psi,$ split $\Psi$ into its even $(\Psi_{even})$ and odd $(\Psi_{odd})$ parts and write $\Psi$ as the 2-vector
\be\label{214} \Psi=\left[\ba{c} \Psi_{even}\\ \Psi_{odd}\ea \right].\ee 
Obviously, 
\be \label{215} R\Psi=\sigma_3 \Psi\ee
in this notation. Moreover, when viewed as acting on wave functions written as in \eref{214} the Hamiltonian \eref{213} takes the form 
\be H=-\frac12 \frac{d^2}{dx^2}+\frac 12(U^2+V^2) +\frac12 \frac{dU}{dx}\sigma_1-\frac12 \frac{dV}{dx}\sigma_3\ee
where 
\be \sigma_1=\left[\ba{cc}0&1\\1&0\ea\right]. \ee

In this formalism, the supersymmetric Hamiltonian associated to the supercharge \eref{211} with R, looks very similar to the one given in \eref{25} especially if $U=0$. It should be stressed however that the standard construction of supersymmetric Hamiltonians, reviewed at the beginning of this section, has nothing to do with the parity properties or parity decomposition of the wave functions. Hence the two supersymmetric realizations (the standard one and the one with reflections) are genuinely different even if they can be, in certain cases, presented in superficially similar forms. 

\section{A Supersymmetric Oscillator with Reflections}
Consider as a first example, the system which is obtained from \eref{211} and \eref{213} by setting 
\bel{31} U=0, \qquad V=x. \ee 
This yields 
\bel{32} Q=\frac1{\sqrt{2}}\left(\frac{d}{dx} R+x\right)\ee
 and 
 \bel{33} H=-\frac12 \frac{d^2}{dx^2}+\frac12 x^2-\frac12R.\ee
This is simply the standard harmonic oscillator to which $(1/2)R$ has been added to render it supersymmetric. This system has been analyzed also in \cite{P3}. It is presented here as illustrative background to the novel supersymmetrization of the Scarf potential that is discussed in the next section. The associated Schr\"odinger equation is readily solved using the familiar orthonormal number states $|n\rangle$, with $n=0,1, \ldots$ and $\langle m|n\rangle=\delta_{m,n}$, of the quantum oscillator. Recall that the annihilation and creation operators $a, \ a^\dagger,$ obeying $[a, a^\dagger]=1$ and realized in the coordinate representation by
\bel{34} a=\frac{1}{\sqrt{2}}\left(\frac{d}{dx}+x\right), \qquad a^\dagger=\frac{1}{\sqrt{2}}\left(-\frac{d}{dx}+x\right)\ee
act as follows on the state $|n\rangle$:
\bel{35} a |n\rangle=\sqrt{n} |n-1\rangle, \qquad a^\dagger |n\rangle =\sqrt{n+1}|n+1\rangle .\ee
The spectrum of $H$ is easily obtained by observing that 
\bel{36}H= a^\dagger a+\frac12 (1-R). \ee
Since 
\bel{37}  R|n\rangle =(-1)^n |n\rangle,\ee
in view of the fact that $\{ R, a\}=\{R, a^\dagger\}=0$ and that we shall take $R|0\rangle=|0\rangle,$
it follows that 
\bel{38} E_n=n+\frac12(1-(-1)^n), \qquad n=0,1,2, \ldots\ee
The spectrum will hence consist only of the even numbers starting with zero. Each level is degenerate except for the ground state which is unique.

It is instructive to diagonalize $Q$. First observe from \eref{34} and \eref{37} that 
\bel{39} Q|n\rangle =\Bigg\{ \ba{cc} \sqrt{n}\  |n-1\rangle & \qquad \mbox{ if n is even,}\\
                                                  \sqrt{n+1} \ |n+1\rangle & \qquad \mbox{ if n is odd.}\ea \ee
In view of \eref{39}, it is readily seen that the states 
\bel{310} |n, \epsilon \rangle =\frac12 \left(|2n+1\rangle+\epsilon |2n+2\rangle\right), \qquad \epsilon =\pm 1\ee
obey 
\bel{311} Q|n, \epsilon \rangle=\epsilon \sqrt{2n+2} |n, \epsilon\rangle,\qquad Q|0\rangle =0 \qquad  n=0,1,\ldots \ . \ee
It thus immediately follows that 
\bel{312} H|n, \epsilon \rangle=(2n+2)|n, \epsilon \rangle,  \qquad H|0\rangle =0 \qquad  n=0,1,\ldots \ee
which is tantamount by linearity to 
\bel{313} H|2n+1\rangle =(2n+2)|2n+1\rangle; \qquad H|2n+2\rangle=(2n+2)|2n+2\rangle. \ee 
 
As is well known, in the coordinate representation, the wave functions $\langle x| n\rangle$ are given in terms of Hermite polynomials $H_n(x)$ by 
\bel{314} \langle x| n\rangle=\frac{1}{\pi^{1/4}2^{n/2}\sqrt{n}}e^{-x^2/2}H_n(x).\ee
Using the relation between the Laguerre polynomials $L_n^\alpha$ and the Hermite polynomials \cite{27}, it is straightforward to find that 
\bel{315} \fl \langle x|n, \epsilon \rangle=\frac{(-1)^n}{\pi^{1/4} }\left[ \frac{n!}{(n+1)_{n+1}}\right]^{1/2}e^{-x^2/2}\left(xL_n^{1/2}(x^2)+\epsilon(n+1)L_{n+1}^{-1/2}(x^2)\right)\ee
where $(a)_n=a(a+1)\cdots(a+n-1)$ is the Pochammer symbol. 

It is readily seen in this example that $R$ maps the degenerate eigenstates into one-another: 
\bel{316} R|n, \epsilon\rangle=-|n, -\epsilon\rangle . \ee
This follows from the fact that in this specific case
\bel{317} \lbrace Q, R \rbrace=0, \qquad [H, R]=0.\ee
Hence, $R$ which was diagonalized simultaneously with $H$, transforms an eigenstate of $Q$ with eigenvalue $\epsilon\sqrt{2n+1}$ into another eigenstate of $Q$, degenerate in energy,  with eigenvalue $-\epsilon\sqrt{2n+1}$. This explains why the levels of the system exhibit a two-fold degeneracy at the exclusion of the ground state. 

\section{ A Novel Supersymmetrization of the Scarf I Potential}
The example of the last section was of course very simple. We shall now present a more elaborate case by providing the supersymmetrization with reflections of the Scarf I Hamiltonian given in \eref{29}. The associated Schr\"odinger equation will be found to be exactly solvable in terms of the recently identified little -1 Jacobi polynomials. 

In the formulation of section 2, let us take 
\bel{41} U(x)=-\frac{\beta}{2\cos x}, \qquad V(x)= -\frac{\alpha}{2 \sin x}.\ee
This choice of functions respect the condition \eref{212}, that is, that $U$ be even and $V$ odd. With these $U$ and $V$, the supercharge \eref{211} and the Hamiltonian \eref{213} read: 
\bel{42} Q_{\alpha, \beta}=\frac{1}{\sqrt{2}}\left( \frac{d}{dx}-\frac{\frac{\beta}{2}}{\cos x}\right) R-\frac{\alpha}{2\sqrt{2}\sin x}\ee
\bea H_{\alpha, \beta}&=&Q^2_{\alpha, \beta}\\
\label{43}&=& -\frac12 \frac{d^2}{dx^2}+\frac{\alpha}{4}\left(\frac{\frac{\alpha}{2}-\cos xR}{\sin^2 x}\right)+\frac{\beta}{4}\left(\frac{\frac{\beta}{2}-\sin x}{\cos^2 x}\right).\eea
This obviously offers an alternative to the standard supersymmetrization \eref{28} of the Scarf I Hamiltonian $H_\beta=H_{0,\beta}$ given in \eref{29}. Note the presence of the reflection operator in $H_{\alpha, \beta}$ when $\alpha\ne 0. $ Two free parameters are present. Observe that $H_\beta$ itself is supersymmetric: $H_\beta=Q^2_{0,\beta}$. We shall now show that the Schr\"odinger equation $H_{\alpha, \beta}\Psi=E\Psi$ is exactly solvable and, to that end, we shall look for the eigenfunctions of $Q_{\alpha, \beta}. $

Let us first remark that in $N=\frac12$ supersymmetry the wave functions are eigenstates of the single supercharge $Q$, they are thus all equivariant and supersymmetry is unbroken even if the ground state does not have zero energy. This is what happens here as it is readily found that the ground state wave function $\Psi_{0; \alpha, \beta}$ is given by 
\bel{44} \Psi_{0; \alpha, \beta}=N_0|\sin x|^{\alpha/2}\cos^{\beta/2}x(1+\sin x)^{1/2}\ee
and satisfies 
\bel{45} Q_{\alpha,\beta}\Psi_{0; \alpha, \beta}=\frac{-1}{2\sqrt{2}}\left(\alpha +\beta+1\right)\Psi_{0; \alpha, \beta}.\ee
The normalization constant $N_0$ is such that 
\bel{460} \int_{\frac\pi 2}^{\frac \pi 2} dx |\Psi_{0; \alpha, \beta}|^2=1.\ee
It is found to be (see \ref{AppendixA}) 
\bel{461} N_0=\left[\frac{\Gamma\left(\frac\alpha 2+\frac\beta 2 +1\right)}{\Gamma\left(\frac \alpha 2+1)\right)\Gamma\left(\frac \beta 2+1)\right)}\right]^{1/2}\ee
with the help of the beta integral. As usual, $\Gamma(x)$ denotes the standard gamma function. 
Let us now carry out the "gauge" transformation of $Q_{\alpha, \beta}$ with the ground state $\Psi_{0; \alpha, \beta}$. Let
\bel{46} \widetilde{Q}_{\alpha, \beta}=\Psi_{0; \alpha, \beta}^{-1} Q_{\alpha, \beta}\Psi_{0; \alpha, \beta}. \ee
It is straightforward to see that 
\bel{47} \widetilde{Q}_{\alpha, \beta}=\frac{\sec x-\tan x}{\sqrt{2}}\frac{d}{dx} -\frac{\alpha\csc x}{2\sqrt{2}}(1-R)-\frac{1}{2\sqrt{2}}(\alpha+\beta+1)R. \ee
Perform now the change of variables \bel{48} y=\sin x\ee to find that 
\bel{49} \widetilde{Q}_{\alpha, \beta}=\frac{1}{\sqrt{2}}(1-y)\frac{d}{dy}R-\frac{\alpha}{2\sqrt{2} y}(1-R)-\frac{1}{2\sqrt{2}}\left(\alpha +\beta+1\right)R. \ee
We thus identify $\widetilde{Q}_{\alpha, \beta}$ as the Dunkl-type operator of which the little -1 Jacobi polynomials are the eigenfunctions. Indeed, it has been shown in \cite{20}, that the little -1 Jacobi polynomials $P_n^{(\alpha, \beta)}(y)$ satisfy the following eigenvalue equation
\bel{410} \fl \left[2(1-y)\frac{d}{dy}R+\left(\alpha+\beta+1-\frac{\alpha}{y}\right)(1-R)\right]P_{n}^{(\alpha, \beta)}(y)=\lambda_{n, \alpha, \beta}P_n^{(\alpha, \beta)}(y)\ee
where 
\bea \label{411} \lambda_{n, \alpha, \beta}=\left\{
 \ba{cc} -2n  &\qquad \mbox{ for $n$ even,}\\
  2(n+\alpha+\beta+1) &\qquad \mbox{ for $n$ odd.}\ea\right. \eea

These polynomials have the following expressions in terms of the hypergeometric (terminating) series: 
\bel{412}\fl P_{n}^{(\alpha, \beta)}(y)=\kappa_n\left[{}_2F_1\left(\ba{cc} -\frac n2 &\frac{n+\alpha +\beta+2}2\\ &\frac{\alpha+1}2 \ea; y^2\right)+\frac{ny}{\alpha+1}{}_2F_1\left(\ba{cc} 1-\frac n2 &\frac{n+\alpha +\beta+2}2\\ &\frac{\alpha+3}2 \ea; y^2\right)\right]\ee
for $n$ even, and 
\bel{413}\fl  P_{n}^{(\alpha, \beta)}(y)=\kappa_n\left[{}_2F_1\left(\ba{cc} \frac{1-n}2 &\frac{n+\alpha +\beta+1}2\\ &\frac{\alpha+1}2 \ea ; y^2\right)-\frac{(\alpha+\beta+1)y}{\alpha+1}{}_2F_1\left(\ba{cc} \frac{1-n}2 &\frac{n+\alpha +\beta+3}2\\ &\frac{\alpha+3}2 \ea; y^2\right)\right]\ee
for $n $ odd. (For a definition of the ${}_2F_1$ symbol see \eref{A11}.) The coefficients $\kappa_n$ are chosen so as to make the polynomials $P_n^{(\alpha, \beta)}(y) $ monic, i.e. $P_n^{(\alpha, \beta)}(y)=y^n+\mathcal{O}(n-1).$ Through the identification of the factor of the leading term in \eref{412} and \eref{413}, they are found to be 
\bel{4140} \kappa_n=\left\{ \ba{cc} 
(-1)^{\frac n2} \frac{\left(\frac {\alpha+1} 2\right)_{\frac n 2 }}{\left(\frac n 2 +\frac \alpha 2 +\frac\beta 2+1\right)_{\frac n 2}}& \qquad \mbox{for $n$ even, } \\
(-1)^{\frac{n+1}{2}} \frac{\left(\frac {\alpha+1} 2\right)_{\frac {n+1} 2 }}{\left(\frac {n+1} 2 +\frac \alpha 2 +\frac\beta 2+1\right)_{\frac {n+1} 2}} & \qquad \mbox{ for $n$ odd. } \ea \right. \ee
For $\alpha > -1$, $\beta >-1$,  they are orthogonal with respect to the weight function 
\bel{414} \omega(y)=|y|^{\alpha}(1-y^2)^{(\beta+1)/2}(1+y). \ee
It thus follows, comparing \eref{49} and \eref{410}, that the wave functions $\Psi_{n; \alpha, \beta}$ defined by 
\bel{415}  \Psi_{n; \alpha, \beta}(x)=\frac{N_n}{N_0} \Psi_{0; \alpha, \beta}P_{n}^{(\alpha, \beta)}(\sin x)\ee
will satisfy the eigenvalue equation 
\bel{416} Q_{\alpha, \beta}\Psi_{n; \alpha, \beta}(x)=q_{n; \alpha, \beta} \Psi_{n; \alpha, \beta}(x)\ee
with 
\be \label{417} q_{n; \alpha, \beta}=\frac{1}{2\sqrt{2}}\Bigg\{ \ba{cc} -(2n+\alpha +\beta+1)& \qquad \mbox{ for $n$ even,}\\ 
 (2n+\alpha +\beta+1)& \qquad \mbox{for  $n$ odd.}\ea \ee
Since $H_{\alpha, \beta}=Q^2_{\alpha, \beta}, $ the spectrum $E_{n; \alpha, \beta}$ of the Hamiltonian is given by 
\bel{418} E_{n;\alpha, \beta}=\frac18 (2n +\alpha+\beta+1)^2, \qquad n=0,1,2,\ldots\ee
and its eigenfunctions are those of $Q_{\alpha, \beta},$ that is, the functions $\Psi_{n; \alpha, \beta}(x)$ given in \eref{415}.

The normalization constants $N_n$ are also chosen so that 
\bel{4200} \int_{-\frac\pi 2} ^{\frac\pi 2} dx |\Psi_{n; \alpha, \beta}(x)|^2=1. \ee
Their calculation, which is described in \ref{AppendixA}, makes use of the moments of the weight function \eref{414} and relies on certain hypergeometric summations. 
They are given by: 
\be \label{Nn} \fl N_n=\left\{ \ba{lc}
 \frac{N_0\left(\frac \alpha 2+\frac \beta 2+1\right)_{n} }{\sqrt{ \left({\frac n2}\right)!\left(\frac \alpha 2 +\frac\beta 2+1\right)_{\frac n2}\left(\frac\alpha 2+\frac 12\right)_{\frac n2}\left(\frac \beta 2+\frac 12\right)_{\frac n2}}} &\quad  \mbox{ for $n$ even,} \\
\frac{N_0\left(\frac \alpha 2+\frac \beta 2+1\right)_{n} }{\sqrt{ \left({\frac n2-\frac12}\right)!\left(\frac \alpha 2 +\frac\beta 2+1\right)_{\frac n2-\frac12}\left(\frac\alpha 2+\frac 12\right)_{\frac n2+\frac12}\left(\frac \beta 2+\frac 12\right)_{\frac n2+\frac12}}},&\quad  \mbox{ for $n$ odd.} \\
\ea \right.\ee

Our experience with the oscillator leads us to examine the action of the reflection operator $R$ on the eigenstates of $Q_{\alpha, \beta}$. The wave functions $R \Psi_{n; \alpha ,\beta}$ will obviously satisfy 
\bel{419} (R Q_{\alpha, \beta}R)R\Psi_{n; \alpha, \beta}=q_{n; \alpha, \beta} R\Psi_{n; \alpha, \beta}, \ee
\bel{420} (R H_{\alpha, \beta}R)R\Psi_{n; \alpha, \beta}=q_{n; \alpha, \beta}^2 R\Psi_{n; \alpha, \beta}. \ee
We may thus couple $H_{\alpha, \beta}$ and $R H_{\alpha, \beta}R$ in a $2 \times 2$ matrix as follows 
\bel{421} \mathcal{H}=\left[\ba{cc}H_{\alpha, \beta} &0\\ 0& RH_{\alpha, \beta}R \ea \right]\ee
to create a system with two-fold degeneracy. As a result, the states 
\bel{422} \left[\ba{c} \Psi_{n; \alpha, \beta}\\ 0 \ea\right], \quad \mbox{and} \quad   \left[\ba{c} 0 \\ R\Psi_{n; \alpha, \beta} \ea\right],\ee 
that are interchanged by the operator 
\bel{423} \left[\ba{cc} 0 & R\\ R & 0 \ea\right],\ee
are degenerate eigenstates of the combined system with $q^2_{n; \alpha, \beta}$ as a common energy. 

Now, it is easy to see that 
\bea \label{424} RQ_{\alpha, \beta} R=-Q_{\alpha, -\beta}\\
\label{425} RH_{\alpha, \beta} R=H_{\alpha, -\beta}.\eea
Hence 
\bel{426} \mathcal{H}=\left[\ba{cc}H_{\alpha, \beta} &0\\ 0& H_{\alpha, -\beta} \ea \right].\ee
When $\alpha=0$, we return to the matrix Hamiltonian \eref{28} that was obtained in the standard way. When $\beta=0$, it is manifest that $H_{\alpha, 0}$ is reflection invariant, $[H_{\alpha, 0}, R]=0$, and has a degenerate spectrum. Indeed, all levels, the ground state included, exhibit a two-fold degeneracy with $\Psi_{n; \alpha, 0}$ and $R\Psi_{n; \alpha, 0}$ satisfying
\bea \label{427} Q_{\alpha, 0}\Psi_{n; \alpha, 0}=\lambda_{n; \alpha, \beta}\Psi_{n; \alpha, 0}\\
\label{428}  Q_{\alpha, 0}R\Psi_{n; \alpha, 0}=-\lambda_{n; \alpha, \beta}R\Psi_{n; \alpha, 0}\eea
and having the same energy. Therefore when $\alpha=0$, we find a situation similar to the one observed for the oscillator except that, here, the reflection symmetry is spontaneously broken. Notwithstanding the properties of the ground state, it is not difficult to convince oneself that such degeneracies will occur whenever $U(x)=0$, that is whenever $[H, R]=0$.

Using the raising and lowering operators of the little -1 Jacobi polynomials, we can obtain intertwining operators that map the eigenfunctions of $Q_{\alpha, \beta}$ into those of $Q_{\alpha, \beta\pm 2}$. Let
%\bel{429o} X_{\alpha, \beta}=\frac{d}{dx}+\frac12 (\beta+1)\tan x-\frac12 \sec x-\frac{\alpha} 2 (1+\csc x)R (old)\ee
\bel{429} X_{\alpha, \beta}=\frac{d}{dx}+\frac12 \beta\tan x-\frac12 \sec x-\frac{\alpha} 2 (1+\csc x)R\ee
 and 
  %\bel{430} Y_{\alpha, \beta}=-\frac{d}{dx}+\frac12 (\alpha+\beta+1)\tan x+\frac{\alpha}{2}\cot x+\frac12 \sec x -\frac{\alpha}{\sin 2x} -\frac{\alpha} 2 (1-\csc x)R,(old)\ee

 \bel{430} Y_{\alpha, \beta}=-\frac{d}{dx}+\frac12 \beta\tan x-\frac12 \sec x -\frac{\alpha} 2 (1-\csc x)R,\ee
 then we have 
 \bea \label{431} X_{\alpha, \beta}\Psi_{n; \alpha, \beta}=[n]_\alpha\frac{N_{n}}{N_{n-1}} \Psi_{n-1; \alpha, \beta+2}\\
  \label{432} Y_{\alpha, \beta}\Psi_{n; \alpha, \beta}=(\beta-1+[n]_\alpha)\frac{N_{n}}{N_{n+1}}  \Psi_{n+1; \alpha, \beta-2}\eea
 where 
 \bel{433} [n]_\alpha=n+\frac{\alpha}2(1-(-1)^n).\ee
 The product of $X_{\alpha, \beta}$ and $Y_{\alpha, \beta}$ is expressible in terms of $H_{\alpha, \beta}$ and $Q_{\alpha, \beta}$ as follows:
 \be Y_{\alpha, \beta+1}X_{\alpha, \beta+1}=2H_{\alpha, \beta}+\sqrt{2}\alpha Q_{\alpha, \beta}+\frac14(\alpha+\beta+1)(\alpha -\beta-1).\ee
 Finally, the operators $X_{\alpha, \beta}$ and $Y_{\alpha, \beta}$ are seen to obey the intertwining relations 
 \bea Q_{\alpha, \beta+2}X_{\alpha, \beta}=-X_{\alpha, \beta}Q_{\alpha, \beta}\\
 Q_{\alpha, \beta-2}Y_{\alpha, \beta}=-Y_{\alpha, \beta}Q_{\alpha, \beta}.\eea
 \section{Conclusion}
 Let us summarize our results to conclude. We considered supersymmetric quantum Hamiltonians that have Dunkl-type operators as supercharges. This approach to supersymmetrization leads to systems that have reflection operators in their Hamiltonians. We introduced in this fashion a supersymmetric extension with two parameters of the Scarf I Hamiltonian in one-dimension. We showed this system to be exactly solvable and found that its wave functions are expressed in terms of the little -1 Jacobi polynomials. 
 
 It would certainly be interesting to further explore models whose supersymmetric extension with reflections would prove exactly solvable. It would also be worth examining how this  approach applies to higher dimensions. 

 \ack
 The authors would like to thank S. Tsujimoto, A. Turbiner and P. Winternitz for discussions. They also wish to thank M. Plyushchay for bringing his related articles to their attention.  The work of (LV) is supported in part through funds provided by the National Science and Engineering Research Council (NSERC) of Canada. (SP) acknowledges a postdoctoral fellowship provided by the Laboratory of Mathematical Physics of the CRM, Universit\'e de Montr\'eal. 

 \appendix
\section{The Normalization of the Wave functions $\Psi_{n; \alpha, \beta}(x)$}\label{AppendixA}
\setcounter{section}{1}
The wave functions of the Hamiltonian \eref{43} are given by 
\bel{A1} \Psi_{n; \alpha, \beta}(x)=\frac{N_n}{N_0} \Psi_{0; \alpha, \beta}(x)P_n^{(\alpha, \beta)}(\sin x)\ee
where 
\bel{A2} \Psi_{0; \alpha, \beta}(x)=N_0 |\sin x|^{\alpha/2} \cos^{\beta/2} x (1+\sin x)^{1/2}.\ee
We shall determine the constants $N_n$ so that 
\bel{A3} \int_{-\frac \pi 2}^{\frac \pi 2} dx |\Psi_{n; \alpha, \beta}(x)|^2=1.\ee
Let $y=\sin x$, 
\begin{eqnarray} \fl \int_{-\frac \pi 2}^{\frac \pi 2} dx |\Psi_{n; \alpha, \beta}(x)|^2&=&
N_n^2\int_{-1}^1\frac{d(\sin x)}{\cos x}|\sin x|^\alpha |\cos x|^\beta (1+\sin
 x)\left(P_n^{(\alpha, \beta)}(\sin x)\right)^2\nn
&=& N_n^2 \int_{-1}^1dy|y|^\alpha (1-y^2)^{\frac{\beta-1}2} (1+y)\left(P_n^{(\alpha, \beta)}(y)\right)^2\nn
&=& N^2_n\int_{-1}^1 dy\  \omega(y)\left(P_n^{(\alpha, \beta)}(y)\right)^2,\end{eqnarray}
where $\omega(y)$ is the measure \eref{414} for which the polynomials $P_n^{(\alpha, \beta)}(y)$ are orthogonal \cite{20}. 

The constant $N_0$ is chosen so that 
\bel{A5}  N_0^2\int_{-1}^1dy \ \omega(y)=1.\ee
It is straightforward to see that 
\bel{A6} \int_{-1}^1dy \ \omega(y) =\int_0^1dt \ t^{\frac{\alpha-1}{2}}(1-t)^{\frac{\beta-1} 2}=B\left(\frac{\alpha}2 +\frac 12, \frac \beta 2 +\frac 12\right)\ee
and hence that 
\bel{A7} N_0=\left[ \frac{\Gamma\left(\frac\alpha 2 +\frac \beta 2 +\frac 12\right)}{\Gamma \left( \frac \alpha 2+\frac 12 \right) \Gamma \left( \frac \beta 2 +\frac12\right)}\right]^{1/2}\ee
where $\Gamma(x)$ and $B (x,y)$ are the standard gamma and beta functions. 
The moments $c_n$, defined by 
\[ c_n=N_0^2 \int_{-1}^1 dy \ \omega(y) y^n\]
are similarly calculated and given by \cite{20}
\bel{A8} c_{2n}=c_{2n-1}=\frac{\left( \frac \alpha 2 +\frac 12\right)_n}{\left(\frac\alpha 2+\frac\beta 2 +1\right)_n}.\ee
Now, from \eref{A3} we have :
\bea \frac{N_n}{N_0}&=&N_0^2\int_{-1}^1 dy \omega(y) \left[P_n^{(\alpha, \beta)}(y)\right]^2\nn
\label{A9}&=& N_0\int_{-1}^1 dy\ \omega(y) P_n^{(\alpha, \beta)}(y) y^n\eea
since the polynomials are monic and obey 
\bel{A10} \int_{-1}^1 dy \omega(y) P_n^{(\alpha, \beta)}(y) y^m=0,\ee
for $m\leq n-1.$ 
Recall that the (generalized) hypergeometric series with $r$ numerator parameters $a_1, \ldots, a_r$ and $s$ denominator parameters $b_1, \ldots, b_s$ are defined by \cite{24, 29}
\bel{A11} {}_rF_s\left(\left.\ba{c} a_1, a_2, \ldots, a_r \\ b_1, \ldots, b_s\ea \right.; z^n\right)=\sum_{n=0}^\infty \frac{(a_1)_n (a_2)_n \cdots (a_r)_n}{(b_1)_n\cdots (b_s)_n} z^n.\ee
In order to evaluate the integral \eref{A9}, one uses the explicit expressions \eref{412} and \eref{413} of the polynomials $P_n^{(\alpha, \beta)}(y)$ in terms of ${}_2F_1$ hypergeometric series and the values of the moments. For $n$ even, $n=2k$, we have 
\bel{A12} \frac{N_0^2}{N_{2k}^2}=\kappa_{2k}\left(A_k+B_k\right)\ee
where 
\bea \label{A13} A_k=\sum_{m=0}^k\frac{(-k)_m\left(k+1+\frac \alpha 2 +\frac \beta 2\right)_m\left(\frac \alpha 2 +\frac 12\right)_{m+k}}
{m!\left(\frac \alpha 2 +\frac12\right)_m \left(\frac \alpha 2 +\frac \beta 2+1\right)_{m+k}}\\
\label{A14} B_k=\frac{2k}{\alpha+1}\sum_{m=0}^{k-1} \frac{(1-k)_m\left(k+1+\frac\alpha 2+\frac \beta 2\right)_m\left(\frac \alpha 2 +\frac 12 \right)_{m+k+1}}
{m!\left(\frac \alpha 2 +\frac32\right)_m \left(\frac \alpha 2 +\frac \beta 2+1\right)_{m+k+1}}\eea
and $\kappa_{2k}$ are the coefficients \eref{4140} ensuring that $P_{2k}^{(\alpha, \beta)}$ is monic. 

Using the identity 
\bel{A15} (a)_{m+k}=(a)_k(a+k)_m, \ee
$A_k$ is reduced to 
\bel{A16} 
A_k=\frac{\left(\frac \alpha 2 +\frac 12\right)_k}{\left(\frac \alpha 2+\frac \beta 2+1\right)_k}\sum_{m=0}^k \frac{(-k)_m\left(k+\frac \alpha 2 +\frac 12\right)_m}{m!\left(\frac \alpha 2+\frac 12\right)_m}\ee
and, with the help of the Chu-Vandermonde summation formula \cite{29} 
\bel{A17} {}_2F_1\left(\left.\ba{cc}-n &, b\\ &c\ea \right. ;1\right)=\frac{(c-b)_n}{(c)_n},\ee
we find 
\bel{A18} A_k=\frac{(-1)^k k!}{\left(\frac \alpha 2 +\frac \beta 2+1\right)_k}.\ee
With the help of \eref{A15} again, $B_k$ can be written as
\bel{A19}\fl B_k=\frac{2k}{\alpha +1}\frac{\left(\frac\alpha 2+\frac 12\right)_{k+1}}{\left(\frac \alpha 2 +\frac \beta 2 +1\right)_{k+1}}
\sum_{m=0}^{k-1}\frac{(1-k)_m\left(k+1+\frac \alpha 2 +\frac \beta 2 \right)_m\left(\frac \alpha 2 +\frac 32 +k\right)_m}{m! \left( \frac \alpha 2 +\frac 32 \right)_m\left(\frac \alpha 2+\frac \beta 2+k+2\right)_m}.\ee
At this point, let us use a summation formula for generalized hypergeometric equations (see for example \cite{30})
\bel{A20}\fl {}_3F_2\left(\left.\ba{ccc} 1-k,& b,& c+k\\ &b+1, & c\ea \right. ;1\right)=\frac{(k-1)!}{(b+1)_{k-1}} \sum_{\ell=0}^{k-1} \frac{b_\ell}{\ell!}{}_2F_1\left(\ba{cc} -\ell, &c+k \\ c \ea ; 1\right)\ee
which is valid for $k\geq 1.$  Using the Chu-Vandermonde summation formula \eref{A17}  twice, we obtain
\bea\fl  {}_3F_2\left(\left.\ba{ccc} 1-k,& b,& c+k\\ &b+1, & c\ea \right. ;1\right)&=&\frac{(k-1)!}{(b+1)_{k-1}} \sum_{\ell=0}^{k-1} \frac{(-k)_k(1-b-\ell)_{\ell}}{\ell! c_\ell}\nn
\fl &=& \frac{(k-1)!}{(b+1)_{k-1}} \sum_{\ell=0}^{k-1} \frac{b_\ell (-k)_k}{\ell!c_\ell}\nn
 &=& \frac{(k-1)!}{(b+1)_{k-1}} \left({}_2F_1\left(\ba{cc} -k, & b\\ & c \ea ; 1\right)-\frac{(-1)^k b_k}{c_k}\right)\nn
&=& \frac{(k-1)!}{(b+1)_{k-1}} \left( \frac{(c-b)_k}{c_k} -\frac{(-1)^k b_k}{c_k}\right)\eea
to see that $B_k,$ for $k\geq 1,$ simplifies to
\bea\label{A24} B_k&=&\frac{(-1)^k k! \left(\frac{\beta}2+\frac12\right)_k}{\left(\frac{\alpha} 2+\frac{\beta}2+1\right)_{2k}} -\frac{(-1)^kk!}{\left(\frac\alpha 2+\frac\beta 2+1\right)_k}\\
&=&\frac{(-1)^k k! \left(\frac{\beta}2+\frac12\right)_k}{\left(\frac{\alpha} 2+\frac{\beta}2+1\right)_{2k}} -A_k.\eea
Note that for $k=0$, $B_k=0$ and so $N_0^2/ N_0^2=A_0=1$ as expected. 
Therefore, for $n$ even, $n=2k, \ k=0,1,\ldots$: 
\bel{A25}\frac{N_0^2}{N_{2k}^2}=\frac{k!\left(\frac\alpha 2+\frac 12\right)_k\left(\frac \beta 2+\frac 12\right)_k\left(\frac \alpha 2 +\frac\beta 2+1\right)_{k}}{\left(\frac \alpha 2+\frac \beta 2+1\right)_{2k}^2}.\ee
For $n$ odd, $n=2k-1,\ k=1,2,\ldots$, one proceeds similarly to find that 
\bel{A26}\frac{N_0^2}{N_{2k-1}^2}=\frac{(k-1)!\left(\frac\alpha 2+\frac 12\right)_k\left(\frac \beta 2+\frac 12\right)_k\left(\frac \alpha 2 +\frac\beta 2+1\right)_{k-1}}{\left(\frac \alpha 2+\frac \beta 2+1\right)_{2k-1}^2}.\ee
This therefore provides the normalization factors $N_n$ as they are given in \eref{Nn}. 

\section{Some (other) examples of Hamiltonians with reflections}\label{AppendixB}

Apart from the little and big -1 Jacobi polynomials, there are other
systems of orthogonal polynomials which satisfy eigenvalue
equations involving Dunkl-type operators. It was shown in  \cite{Cheikh} 
that the generalized Hermite and generalized Gegenbauer
polynomials are the only symmetric orthogonal polynomials that obey such an equation, in these cases of second order with respect to
the classical Dunkl operator. We indicate here that the equation for the generalized Gegenbauer
polynomials can be presented in  Schr\"odinger form
with an additional "reflection" term.

The generalized Gegenbauer polynomials \cite{Chih, Belm}
$P_n(y)$ are symmetric  polynomials (i.e. $P_n(-y) =
(-1)^n P_n(y)$) which are orthogonal on the interval $[-1,1]$ with
respect to the weight function \be w(x) = |x|^{2 \mu}
(1-x^2)^{\alpha}. \lab{w_Geg}\ee The polynomials $P_n(y)$ satisfy the eigenvalue equation \cite{Cheikh}
 \be L P_n(y) = \lambda_n P_n(y), \lab{eig_G} \ee
 where 
\be L= (1-y^2) T_{\mu}^2 -2(\alpha+1) y T_{\mu} \lab{L_G} \ee 
and $T_{\mu}$ is
the classical Dunkl operator
\be
T_{\mu} = \partial_y + \mu y^{-1} (I-R).
\ee
The eigenvalues are 
\be \lambda_n = \left\{ {-n(n+1+2\alpha+2\mu) 
\quad  \mbox{ for $n$ even,}    \atop -(2\mu+n)(2\alpha+n+1) \quad
\quad \mbox{for $n$ odd.}} \right .  \lab{lam_G} \ee 
Change the independent variable $y=\sin x$ and consider the operator 
\be H= -F_0(x) L F_0^{-1}(x), \lab{HG} \ee
 where \be F_0(x) = \sqrt{w(y) \cos x} =
|\sin x|^{\mu} \cos^{\alpha+1/2} x. \lab{F0} \ee 
It is assumed that $-\pi/2<x<\pi/2$. It can be checked that the operator $H$ has the form 
\be H= -\partial_x^2 + U_0(x) + U_1(x) R, \lab{H_G} \ee where \be U_0(x)
= \frac{\alpha^2 \cos^4 x +(\mu^2 -2 \alpha^2 +1/4) \cos^2 x +
\alpha^2-1/4}{\cos^2 x \sin^2 x} -\alpha \lab{U0} \ee and \be
U_1(x) = (2 \alpha+1) \mu - \frac{\mu}{\sin^2 x}. \lab{U1} \ee
 This is another example of exactly solvable Schr\"odinger Hamiltonian which includes the reflection operator $R$.

Note that the function $F_0(x)$ defined in \eref{F0} is the ground state wave function of the Hamiltonian $H$ corresponding to the lowest
eigenvalue $\lambda_0=0$: \be H F_0(x) =0.\ee The bound state wave functions $\psi_n(x)$ that satisfy the Schr\"odinger equation
\be  H \psi_n(x) = \lambda_n \psi_n(x) \ee
have the form
 \be \psi_n(x) = F_0(x) P_n(\sin x),
\lab{psi_n} \ee where $P_n(y)$ are generalized Gegenbauer
polynomials.

There are two special cases of the Hamiltonian \eref{H_G}
worth mentioning. If $\mu=0,$ then the term $U_1(x)R$ with the reflection operator 
disappears and the Hamiltonian $H$ becomes the usual trigonometric
P\"oschl-Teller trigonometric potential: \be H= -\partial_x^2 +
\frac{\alpha^2-1/4}{\cos^2 x} - \frac{(2 \alpha+1)^2}{4}.
\lab{H_PT} \ee It is well known that its eigenfunctions are
expressed in terms of ordinary Gegenbauer (ultraspherical)
polynomials.

Another interesting special case occurs for $\alpha=-1/2$. We have then  \be H = -\partial_x^2 + \frac{\mu^2}{\sin^2 x} -
\mu^2 -\frac{\mu}{\sin^2 x} R \lab{H_S} \ee which can be related to the two-particle
Calogero-Sutherland-Moser (CSM) model with an exchange
term. The N-body CSM Hamiltonian (with the exchange operators) is 
\cite{BDF} \be H=-\sum_{j=1}^N \frac{\partial^2}{\partial x_j^2} +
\beta \gamma^2 \sum_{j,k=1 \atop j<k}^{N}
\frac{\beta/2-S_{jk}}{\sin^2[\gamma (x_j-x_k)]}, \lab{Suth_E} \ee
where $\beta, \gamma$ are arbitrary real parameters and $S_{jk}$
is the operator which exchanges the coordinates.

Let N=2, put
$\gamma=2^{-1/2}, \: \beta=2 \mu$ and choose the coordinates
\be 
x=\frac{x_1-x_2}{\sqrt{2}}, \quad u = \frac{x_1+x_2}{\sqrt{2}}.
\ee
We can then rewrite the Hamiltonian \eref{Suth_E} as
 \be H =
-\partial_x^2 - \partial_u^2 + \frac{\mu^2}{ \sin^2 x} -
\frac{\mu}{ \sin^2 x}R .\lab{Suth_xu} \ee The term $-
\partial_u^2$ corresponds to the conserved energy of the center-of-mass and can be separated out. Comparing \eref{Suth_xu} with \eref{H_S} we
see that these two Hamiltonians coincide up to an inessential
constant term.

Another one-dimensional quantum Hamiltonian with a reflection term can similarly be obtained from the two-particle rational CMS model; the wave functions in this case are expressed in terms of generalized Hermite polynomials.


\section*{References}

\begin{thebibliography}{10}

\bibitem{1} Witten E  1981 Dynamical breaking of supersymmetry \emph{  Nucl, Phys B} {\bf 188}  513-554

\bibitem{2}  Cooper F, Khare A and Sukhatme U 2001 \emph{ Supersymmetry in Quantum Mechanics} (Singapore: World Scientific)


\bibitem{P1} Plyushchay M S 1994 Supersymmetry without fermions DFTUZ-94-05, hep-th/9404081 

\bibitem{P2} Plyushchay M S 1996 Deformed Heisenberg algebra, fractional spin fields and supersymmetry without fermions \emph{ Annals Phys.} {\bf 245} 339-360 


\bibitem{3}  Calogero F 1969 Solutions of a three-body problem in one-dimension \emph{ J. Math. Phys. } {\bf 10} 2191-2196

\bibitem{4} Sutherland B 1971 Quantum many-body problem in one-dimension, I, II, \emph{ J. Math. Phys.}  {\bf 12}  246-250 

\bibitem{5} Minahan J A and  Polychronakos A P 1993 Integrable systems for particles with internal degrees of freedom \emph{ Phys. Lett. B }{\bf 302}  265-270

\bibitem{6} Polychronakos A P 1992 Exchange operator formalism for integrable systems of particles \emph{ Phys. Rev. Lett. } {\bf 69} 703-705 

\bibitem{7} Brink L, Hansson T H, Konstein S and Vasiliev M A 1993 The Calogero model-anyonic representations, fermionic extension and supersymmetry\emph{ Nucl. Phys. B }{\bf 401} 591-612

\bibitem{8} Lapointe L and Vinet L 1996 Exact operator solutions of the Calogero-Sutherland model  \emph{ Comm. Math. Phys. }{\bf 178}  425-452


\bibitem{BDF} Baker T H, Dunkl C F and Forrester P J 2000 Polynomial eigenfunctions of the Calogero - Sutherland - Moser models with exchange terms \emph{ Calogero - Sutherland - Moser Models CRM Series in Mathematical Physics} ed J F van Diejen and L Vinet (New York: Springer) pp 37-51

\bibitem{FLV} Floreanini R,  Lapointe L,  and Vinet L 1996 The polynomial SU(2) symmetry algebra of the two-body Calogero model\emph{ Phys. Lett. B} {\bf 389} 327-333 

\bibitem{9} Dunkl C F 1989 Differential-difference operators associated to reflection groups \emph{ Trans. Amer. Math. Soc.} {\bf 311}  167-183

\bibitem{10} Sharma J K, Mehta C L and  Sudarshan E C G 1978 Para-Bose coherent states \emph{J. Math. Phys.} {\bf 19} 2089-2093

\bibitem{11}  Sharma J K ,   Mehta C L, Mukunda N and Sudarshan E C G 1981 Representations and properties of para-Bose oscillator. II Coherent states and the minimum uncertainty states \emph{ J. Math. Phys.} {\bf 22} 78-90

\bibitem{12} Macfarlane A J 1994 Generalized oscillator systems and their parabosonic interpretation \emph{ Proc. Int. Workshop on Symmetry Methods in Physics (JINR, Dubna, 1994)} ed A N Sissakian, G S Pogosyan and S I Vinitsky p 319

\bibitem{13} Quesne C 2010 Superintegrability of the Tremblay-Turbiner-Winternitz quantum quantum Hamiltonians on a plane for odd $k$ \emph{ J. Phys. A} {\bf 43}  082001

\bibitem{14} Tremblay F, Turbiner A V and Winternitz P 2009 An infinite family of solvable and integrable quantum systems on the plane\emph{ J. Phys. A} {\bf 42}  242001

\bibitem{15} Jafarov E I, Stoilova N I and van der Jeugt J 2011 Finite oscillator models: The Hahn oscillator \emph{ J. Phys. A } {\bf 44} 265203

\bibitem{16} Tsujimoto S, Vinet L and  Zhedanov A 2011
From $sl_q(2)$ to a parabosonic Hopf algebra arXiv 1108.1603

\bibitem{17} Kay A 2010 A review of perfect state transfer and its applications as a constructive tool\emph{ Int. J. Quantum Inf.} {\bf 8} 641

\bibitem{18}     Dunkl C F and Xu Y 2001 Orthogonal polynomials of several variables \emph{Encyclopedia of Mathematics and Its Applications} vol 81 (Cambridge: Cambridge University Press)

\bibitem{19} R\"osler M 2003 Dunkl operators: theory and applications \emph{ Orthogonal Polynomials and Special Functions (Lecture Notes in Mathematics)} vol 1817 (Berlin: Springer) pp 93-135 

\bibitem{20} Vinet L and  Zhedanov A 2011 A "missing" family of classical orthogonal polynomials \emph{ J. Phys. A }{\bf 44} 085201

\bibitem{21}  Vinet L and  Zhedanov A 2011 A limit $q\rightarrow -1$ for big q-Jacobi polynomials\emph{ Trans. Amer. Math. Soc} (to appear) arXiv: 1011.1429v3

\bibitem{22} Vinet L and  Zhedanov A 2011 A Bochner theorem for Dunkl polynomials, SIGMA {\bf 7}  020

\bibitem{23} Tsujimoto S, Vinet L and  Zhedanov A 2011 Jordan algebra and orthogonal polynomials \emph{ in preparation}

\bibitem{24} Koekoek R, Lesky P and Swarttouw R 2010 Hypergeometric Orthogonal Polynomials and Their Q-Analogues (Berlin: Springer-Verlag)

\bibitem{25} Scarf F 1958  New solvable energy band problem \emph{ Phys. Rev. }{\bf 112}  1137-1140

\bibitem{Chih} Chihara T 1978 {\it An Introduction to Orthogonal
Polynomials} (New York: Gordon and Breach) 

\bibitem{Belm} Belmehdi S 2001 Generalized Gegenbauer polynomials \emph{
J. Comput. Appl Math. } {\bf 133}  195--205

\bibitem{Cheikh}  Ben Cheikh Y and Gaied M 2007 Characterization of
the Dunkl-classical symmetric orthogonal polynomials \emph{ Appl. Math.
and Comput. } {\bf 187} 105--114


\bibitem{26} D'Hoker E and Vinet L 1985 Dynamical supersymmetry of the magnetic monopole and the $1/r^2$-potential \emph{ Comm. Math. Phys. }{\bf 97} 391-427

\bibitem{Quesne2009} Quesne C 2009 Solvable rational polynomials and exceptional polynomials in supersymmetric quantum mechanics \emph{SIGMA} {\bf 5} 084


\bibitem{P3} Gamboa J, Plyushchay M S, and Zanelli J 1999 Three aspects of bosonized supersymmetry and linear differential field
equation with reflection \emph{
Nucl.Phys.B} {\bf 543} 447-465



\bibitem{27}  Erd\'elyi A (ed) 1958 \emph{ Higher Transcendental Functions} vol II (New York: McGraw-Hill)

\bibitem{29} Gasper G and Rahman M 1990  Basic Hypergeometric Series \emph{ Encyclopedia of Mathematics and Its Applications} vol 35 (Cambridge: Cambridge University Press) 

\bibitem{30} Prudnikov A P, Brychkov Y A and Marichev O I 1988 \emph{ Integrals and Series, Volume 2: Special Functions} (New York: Gordon and Breach) 










\end{thebibliography}
\end{document}